\documentclass[12pt]{JHEP3}

\usepackage{amsmath,amssymb,axodraw}

\newcommand\fverb{\setbox\pippobox=\hbox\bgroup\verb}
\newcommand\fverbdo{\egroup\medskip\noindent%
                        \fbox{\unhbox\pippobox}\ }
\newcommand\fverbit{\egroup\item[\fbox{\unhbox\pippobox}]}
\newbox\pippobox

\title{\Large\bf Quantum numbers of heavy
neutrinos, tri-bi-maximal mixing through double seesaw with
permutation symmetry, and comment on $\theta_{\rm
sol}+\theta_c\simeq \frac{\pi}{4}$}

\preprint{SNUTP 05-019}

\author{
Jihn E. Kim  and Jong-Chul Park
  \\
  School of Physics and Center for Theoretical Physics,
  Seoul National University, Seoul 151-747, Korea \\
  E-mail: \email{
  jekim@phyp.snu.ac.kr,\ jcpark@phya.snu.ac.kr} }

\maketitle

\abstract{Using the family symmetry, in the neutrino mass matrix we
remove the Yukawa coupling (arising in the Dirac type mass between
the heavy neutrinos and light lepton doublets) dependence in the
double seesaw mechanism so that it is directly proportional to the
mass matrix $m^{(nn)}$ of heavy Majorana neutrinos. The family
symmetry is supposed to be broken spontaneously at high energy scale
so that the neutrino mass matrix is given by the family symmetry at
high energy scale. With the permutation symmetry $S_3$, we note a
variety of possible mass hierarchies arising distinctly in
neutrinos, charged leptons, $Q_{\rm em}=-\frac13$ quarks, and
$Q_{\rm em}=\frac23$ quarks. Distinguishing these hierarchies, we
obtain a relation between the CKM angles and the MNS angles.
Finally, we comment on the approximate relation $\theta_{\rm
sol}+\theta_c\simeq \frac{\pi}{4}$.}


\begin{document}

\def\lsl{ l \hspace{-0.45 em}/}
\def\ksl{ k \hspace{-0.45 em}/}
\def\qsl{ q \hspace{-0.45 em}/}
\def\psl{ p \hspace{-0.45 em}/}
\def\ppsl{ p' \hspace{-0.70 em}/}
\def\dsl{ \partial \hspace{-0.45 em}/}
\def\Dsl{ D \hspace{-0.55 em}/}
\def\matrix{ \left(\begin{array} \end{array} \right) }
\def\frsqotw{\frac{1}{\sqrt2}}
\def\frsqoth{\frac{1}{\sqrt3}}
\def\frsqtth{\sqrt{\frac{2}{3}}}
\def\frsqo{\frac{\omega}{\sqrt{3}}}
\def\frsqoo{\frac{\omega^2}{\sqrt{3}}}

\def\hf{\textstyle{\frac12~}}
\def\hff{\textstyle{\frac13~}}
\def\hfg{\textstyle{\frac23~}}
\def\DQ{$\Delta Q$}

\def\Qem{Q_{\rm em}}

\thispagestyle{empty}

\baselineskip 0.7cm

%%%%%%%%%%%%%%%%%%%%%%%%%%%%%%%%%%%%%%%%%%%%%%%%%%%%%%%%%%%%%%%%%%%%
%%%%%%%%%%%%%%%%%%%%%%%%%%%%%%%%%%%%%%%%%%%%%%%%%%%%%%%%%%%%%%%%%%%%
\section{Motivation}

Neutrino oscillations are parametrized by the MNS unitary matrix
\begin{equation}
U_{\rm MNS} =\left(\begin{array}{ccc}
U_{e1}& U_{e2}&U_{e3}\\[0.2em]
U_{\mu 1}&U_{\mu 2}&U_{\mu 3}\\[0.2em]
U_{\tau 1}&U_{\tau 2}&U_{\tau 3}
\end{array}\right)\label{MNSdef}
\end{equation}
where $\alpha=\{e,\mu,\tau\}$ is the weak eigenstate index and
$i=\{1,2,3\}$ is the mass eigenstate index. The
$|\alpha\rangle\to|\beta\rangle$ transition amplitude in the time
interval $t$ is \cite{FYbook},
\begin{equation}
\langle\nu_\beta|\nu_\alpha\rangle_t=\sum_{j} U_{\alpha j}
U^\dagger_{j\beta}e^{-iE_jt}.
 \end{equation}
Then, the survival probability of flavor $\nu_\alpha$ at high energy
$E$ is given by
\begin{equation}
P_{\nu_\alpha\to\nu_\alpha}=1-\sum_{i,j}4|U_{\alpha i}|^2|U_{\alpha
j}|^2\sin^2\textstyle\left(\frac{\Delta m^2_{ij}}{4E}t\right)
\end{equation}
where $\Delta m^2_{ij}=m^2_i-m^2_j$. Currently, the disappearance
data from atmospheric and solar neutrinos point toward the following
form,
\begin{equation}
U_{\rm MNS}=\left(
\begin{array}{ccc}
 \frsqtth e^{i\delta_3}\quad&\times \quad&0 \\[0.2em]
\times\quad &\times \quad&\frsqotw e^{i\delta_2}\\[0.2em]
\times\quad &\times\quad &\frsqotw e^{i\delta_1}
\end{array}\right)
\end{equation}
where $\times$ is unspecified. Motivated by this observation,
recently a {\it tri-bi-maximal} mixing form has been suggested
\cite{Wolfenstein,HPS02,Ma1},
\begin{equation}\label{tribi}
U_{\rm MNS}\simeq\left(\begin{array}{ccc}
 \frsqtth \quad&\frsqoth\quad &0 \\[0.2em]
 -\frac{1}{\sqrt6}\quad&\frsqoth\quad &\frsqotw\\[0.2em]
-\frac{1}{\sqrt6}\quad&\frsqoth\quad &-\frsqotw
\end{array}\right)
\end{equation}
which is possible starting from some discrete symmetries such as the
permutation symmetry $S_3$ \cite{HPS02} and tetrahedral symmetry
$A_4$ \cite{Ma1}. Note that there does not exist  a measurable CP
phase in this form. Since the neutrino mixing matrix involves the
unitary matrices diagonalizing  charged lepton and  neutrino mass
matrices, one must consider both of these unitary matrices. With the
$S_3$ symmetry, for example, the charged lepton  and neutrino mass
matrices are assumed to take different representations under $S_3$.
One simple choice is assuming that charged leptons are singlets
under the discrete group and the form (\ref{tribi}) is obtained
purely from the neutrino mass matrix. Another possibility is to
assume a bi-maximal form for the neutrino mass and a tri-maximal
form for the charged lepton mass as done in Ref. \cite{HPS02}.
Certainly, the latter choice is very appealing in the simplicity of
explaining both the bi-maximal \cite{bi} and tri-maximal
\cite{HPS02} structures in a single mixing matrix of (\ref{tribi}).

But there exists another complication due to the hypothetical
mechanism for generating neutrino masses. In the standard model(SM),
there exist renormalizable couplings for charged lepton
masses.\footnote{Renormalizable and nonrenormalizable couplings are
the effective ones at low energy.} But to generate neutrino masses
at the SM level, non-renormalizable dimension-5 couplings are
needed. To obtain these through the seesaw mechanism, one needs
heavier neutrinos, collectively represented as $n$. Thus, the above
attractive proposal for the bi- and tri- structure has to be
carefully addressed.

It is of utmost importance to relate the neutrino mass matrix and
the $n$ mass matrix, or there are too many parameters to be assumed
to specific values. In the seesaw scenario, there appear Yukawa
couplings between singlet neutrinos and doublet neutrinos, which
complicate a direct application of symmetry idea. In this regard,
earlier Lindner et al. \cite{Lindner} studied the possibility of
removing this Yukawa coupling dependence as `screening of Dirac
flavor structure'.

For this purpose, we introduce a family symmetry and use the double
seesaw mechanism to relate neutrino and $n$ mass matrices. In this
process, we need two types of heavy neutrinos, collectively
represented as $n$ and $N$ types and two continuous symmetries $F_1$
and $F_2$. Specifically, the dependence of neutrino mass matrix on
the Yukawa couplings involving $N$ and lepton doublets are removed,
which will be shown to be possible by a hierarchy of singlet vacuum
expectation values(VEVs).

Another appealing phenomenological relation is the sum rule of the
solar neutrino mixing, $\theta_{\rm sol}$, and the Cabibbo angle,
$\theta_c$,
\begin{equation}
 \theta_{\rm sol}^{\rm exp}+\theta_c^{\rm exp}
 \simeq 33^{\rm o}+13^{\rm o}\to\frac{\pi}{4}.
 \label{sumatca}
\end{equation}
Already there exist many ideas trying to explain the above relation
\cite{bi,complem} with GUTs and with the quark-lepton
complementarity idea,\footnote{Some think that the quark-lepton
complementary relation is just a numerical accident
\cite{XingJarl}.} but most of them do not remove the Yukawa coupling
dependence in the neutrino mass matrix. We find that Ref.
\cite{Lindner} independently observes the same kind of the removal
of Yukawa coupling dependence under the phrase, $\lq$screening of
Dirac flavor structure'. In Eq. (\ref{sumatca}), $\theta_{\rm sol}$
appears in the MNS matrix which is given by diagonalizing neutrino
and charged lepton mass matrices,
\begin{equation}
 U_{\rm MNS}=U^\dagger_l U_\nu,\label{UMNS}
\end{equation}
and $\theta_c$ appears in the CKM matrix which is obtained by
diagonalizing $Q=\frac23$ and $Q=-\frac13$ quark mass matrices,
\begin{equation}
U_{\rm CKM}=U^\dagger_u U_d.\label{UCKM}
\end{equation}
 To relate the mixing angles of the leptonic sector and the
quark sector, one must unify leptons and quarks, or go beyond the SM
to grand unified theories(GUTs) or the quark-lepton complementarity.
Here, we will be interested in the above sum rule also, and employ
the quark-lepton complementarity  idea. But we will not discuss any
specific model in detail.

At the SM level, there are four types of mass matrices:
$\Qem=-\frac13$ quark mass matrix $m^{(d)}$, $\Qem=\frac23$ quark
mass matrix $m^{(u)}$, $\Qem=-1$ charged lepton mass matrix $m_l$,
and $\Qem=0$ neutrino mass matrix $m_\nu$. GUTs relate some of these
mass matrices. The well-known one is the SU(5) relation with a Higgs
quintet generating both $\Qem=-\frac13$ quark and $\Qem=-1$ lepton
mass matrices \cite{BEGN}. Then, we obtain a relation between four
unitary matrices, $U_{\rm CKM}, U_{\rm MNS}$, and two unitary
matrices $U_\nu$ and $U^{(u)}$ which diagonalize $m_\nu$ and
$m^{(u)}$, respectively. Here, usually $U_{\rm CKM}$ and $U_{\rm
MNS}$ are phenomenologically determined and $U_\nu$ and $U^{(u)}$
are given theoretically. Thus, in addressing the above questions, it
is suggested that the relation arises naturally from a proposed
discrete symmetry.

The four types of fermions have distinct mass hierarchies. The
charged leptons and down type quarks have the similar pattern for
masses, $m_{e,d}\ll m_{\mu,s}\sim \frac{1}{20}m_{\tau,b}$. The
neutrino mass hierarchy is quite different from this,
\begin{equation}
\Delta m^2_{\nu\ ij}\ll \Delta m^2_{\nu\ jk}\label{nupattern}
\end{equation}
where the LHS is for the solar neutrino oscillation and the RHS is
for the atmospheric neutrino oscillation. Finally, the up type quark
masses are distinct from any of the above patterns,
\begin{equation}
m_u\ll m_c\sim \frac{1}{150}m_t,\quad {\rm or}\quad m_u,m_c\ll m_t.
\label{uppattern}
\end{equation}
The pattern (\ref{nupattern}) hints two almost degenerate neutrinos
compared to the other neutrino, and the pattern (\ref{uppattern})
hints two almost massless quarks compared to top quark. These
observations can be used as an input in the mixing angle relation.

In the SM without a family symmetry, there results the CKM mixing
since the Yukawa couplings for the up-type quark masses are given
differently from those for the down-type quark masses. With a family
symmetry, nonzero mixing angles can arise only after breaking the
imposed family symmetry. Since low energy Yukawa couplings are given
in terms of dimension 4 renormalizable couplings, we assume that the
original high energy couplings are non-renormalizable. If we
consider an $S_3$ family symmetry, the quark mixing can arise only
after the $S_3$ symmetry is spontaneously broken differently for the
up-type and down-type quark sectors. The same argument applies to
the leptonic sectors also.

 In Sec. \ref{sec:nunrel}, we introduce two
continuous quantum numbers $F_1$ and $F_2$ to relate the neutrino
masses and the $n$ masses via the double seesaw mechanism. In Sec.
\ref{sec:MixingLH}, distinct patterns of neutrinos, charged leptons,
$\Qem=-\frac13$ and $\Qem=\frac23$ masses are used to obtain the
tri-bi-maximal MNS unitary matrix.  In Sec. \ref{sec:CKMMNSrel}, we
relate $U_{\rm CKM}$ and $U_{\rm MNS}$ and obtain an approximate
relation $\theta_{\rm sol}+\theta_c\simeq\frac{\pi}{4}$. Sec.
\ref{sec:Conclusion} is a conclusion.

%%%%%%%%%%%%%%%%%%%%%%%%%%%%%%%%%%%%%%%%%%%%%%%%%%%%%%%%%%%%%%%%%%%%%%%%%%
%%%%%%%%%%%%%%%%%%%%%%%%%%%%%%%%%%%%%%%%%%%%%%%%%%%%%%%%%%%%%%%%%%%%%%%%%%
\section{Neutrino masses induced by heavy neutrinos}\label{sec:nunrel}

The charged lepton and quark masses arise from the dimension four
Yukawa couplings
\begin{equation}
-{\cal L}_Y=f^{(u)}_{IJ}u^{c I}H_2q^J +f^{(d)}_{IJ}d^{c
I}H_1q^J+f^{(e)}_{IJ}e^{c I}H_1l^J+{\rm h.c.} \label{qlYukawa}
\end{equation}
where all the fermions are represented in terms of left-handed Weyl
fields, $q$ and $l$ are the quark and lepton doublets, and we used
the two Higgs doublet notation with hypercharges $Y(H_1)=-\frac12$
and $Y(H_2)=\frac12$. If we introduce only one Higgs doublet $H_1$,
then we replace $H_2$ by $-i\sigma_2 H_1^*$. In Eq.
(\ref{qlYukawa}), the  roman characters $I,J$ represent the family
indices. The quark and lepton masses between families are
distinguished by the difference of their Yukawa coupling strengths.
The smallness of Cabibbo angle in the two family case is due to the
hierarchy $f_{12},f_{21},f_{11}\ll f_{22}$. For the three family
case, we have $f_{22},f_{i3}(i=1,2)\ll f_{33}$.

\begin{figure}[h]
\begin{center}
\begin{picture}(400,170)(0,0)
\Line(80,80)(300,80)

\SetWidth{1.2} \Text(75,80)[r]{$\nu_I$}
\DashLine(110,80)(110,130){5}\Text(110,140)[c]{$vf_{IK}$}
\DashLine(150,80)(150,30){5}\Text(150,20)[c]{$(M^{-1})_{KP}$}
\DashLine(190,80)(190,130){5}\Text(190,140)[c]{$m_{PQ}$}
\DashLine(230,80)(230,30){5}\Text(230,20)[c]{$(M^{-1})_{QL}$}
\DashLine(270,80)(270,130){5}\Text(270,140)[c]{$vf_{LJ}$}
\Text(305,80)[l]{$\nu_J$}

\Text(130,90)[c]{$N_{K}$}\Text(250,90)[c]{$N_{L}$}
\Text(170,90)[c]{$n_{P}$}\Text(210,90)[c]{$n_{Q}$}
\end{picture}
\caption{A double seesaw diagram with generic eigenvalues of $fv,m$
and $M$ have a hierarchy $M\gg m\gg fv$.}\label{seesawdoub}
\end{center}
\end{figure}
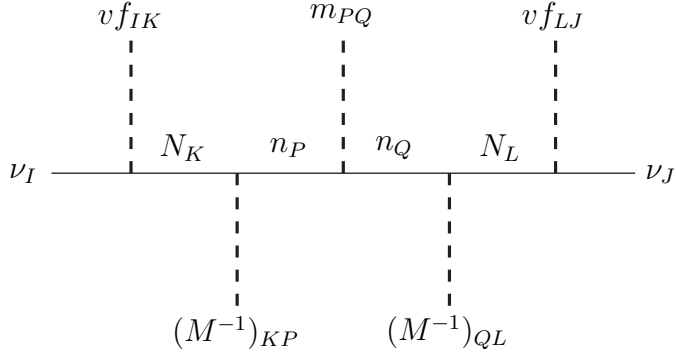

The gauge symmetry does not allow masses of neutrinos. To obtain
neutrino mass, we must introduce more field(s). We adopt the seesaw
idea of introducing SM singlet (neutral) heavy fermion(s), $n$. Then
neutrino masses can arise through the seesaw mechanism, symbolically
written as
\begin{equation}
m_\nu\sim \frac{(fv)^2}{\tilde M}\label{seesawsing}
\end{equation}
where $v$ is the vacuum expectation value of the Higgs doublet
$H_2$, and ${\tilde M}$ is the Majorana mass of the heavy
neutrino(s). Because the Yukawa coupling appears as $f^2$ in the
numerator, it is not expected that the single seesaw would remove
the $f^2$ dependence. To remove the $f^2$ dependence, we must have
the same $f^2$ appearing in the denominator also. For this purpose,
a double seesaw is needed as depicted in Fig. \ref{seesawdoub}.
In another context, the double seesaw was considered in Ref.
\cite{doubsee}, where however our attempt of removing the Yukawa
coupling dependence was not tried. Also, a kind of $U(2)$ symmetry
for dimension-5 neutrino mass operator was considered
\cite{RossRaby}, which does not belong to our scheme either. To
relate Yukawa couplings appearing in the Dirac masses $fv$ and $M$,
we must introduce some symmetry. So let us introduce family quantum
number $Q_F$. For each family, let us introduce the following chiral
fermions
\begin{align}
l_I\equiv \left(\begin{array}{c}\nu_I\\ l_I\end{array}\right),\ \
l^c_I,\ \ N_I,\ \ n_I,\ \ q_I,\ \ u^c_I,\ \ d^c_I
\end{align}
where $N_I$ and $n_I$ are neutral SU(2)$\times$U(1)$_Y$ singlet
heavy neutrinos and $l$ and $q$ are the lepton and quark doublets of
the SM. One family is composed of 17 chiral fields, which together
with Higgs multiplets can arise from the $E_6$ GUT with {\bf 27}
\cite{Esix} and trinification with
$\bf(3,3^*,1)+(1,3,3^*)+(3^*,1,3)$ \cite{kimjheptrit,trikim04}.

The SM singlet neutral leptons can have bare masses unless they are
forbidden by a symmetry. The lepton number is a good symmetry
forbidding their bare masses. We assign the opposite lepton numbers
to $l_I$ and  $N_I$. For $n_I$, we introduce another independent
quantum number, say $n$-number. $l_I$ and $N_I$ do not carry the $n$
number. The symmetry of leptons is
SU(2)$_L\times$U(1)$_Y\times$U(1)$_{F_1}\times$U(1)$_{F_2}$ where
U(1)$_{F_1}\times$U(1)$_{F_2}$  is a continuous symmetry. To
generate fermion masses, let us introduce the usual
SU(2)$\times$U(1)$_Y$ doublet Higgs field $H_2$ and
SU(2)$\times$U(1)$_Y$ singlet but U(1)$_{F_2}$ nonsinglet $S$s. The
lepton quantum numbers, $F_1$ and $F_2$, are
\begin{equation}\label{QNs}
\begin{array}{cccccccc}
& l\quad& e^c\quad& N\quad & n\quad & S_1\quad& S_2\quad&H_2\\
F_1\quad &1\quad&-1\quad&-1\quad&1\quad&-2\quad&0\quad&0\\
F_2\quad&0\quad&0\quad&0\quad&1\quad&-2\quad&-1\quad&0
\end{array}
\end{equation}
Consistently with the quantum numbers of (\ref{QNs}), we can write
the renormalizable Yukawa couplings involving singlet leptons as
\begin{equation}
-{\cal L}=f^{(lN)}_{IJ}N^{I}H_2
l^J+f^{(Nn)}_{IJ}N^{I}n^{J}S_2+f^{(nn)}_{IJ}n^{I}n^{J}S_1+{\rm
h.c.} \label{NnYukawa}
\end{equation}
For a family symmetry, we require that $f_{IJ}$ are the same if
$l,e^c,N$ and $n$ belong to the same family, i.e.
\begin{align}
&f^{(lN)}_{IJ}=f^{(Nn)}_{IJ}\to\label{FamSym}\\
&?=f^{(nn)}_{IJ}.\label{FamSymMaj}
\end{align}
The relation (\ref{FamSym}) for complex Dirac masses can be
differentiated in principle from the relation (\ref{FamSymMaj}) for
real Majorana masses. But the family symmetry can be achieved by
assigning $l$ and $n$ in the same multiplet. Note that the $F_2$
quantum numbers of $l$ and $n$ in (\ref{QNs}) are different; thus we
interpret $F_2$ as a $U(1)$ subgroup of a unifying group so that $l$
and $n$ can be put in the same representation of the unifying group.

The double seesaw diagram of Fig. \ref{seesawdoub} gives neutrino
masses. We can see immediately that for $f^2$ to appear in the
denominator, $M$ of Fig. \ref{seesawdoub} can be taken to be much
larger than those of $m$. However, it is known that even without
this restriction the Yukawa coupling dependence disappears
\cite{Lindner}. However, an intuitive understanding of this
phenomenon is most transparent in the limit
\begin{equation}
V_2\gg V_1 \gg v,\label{hiera}
\end{equation}
where $V_1=\langle S_1\rangle, V_2=\langle S_2\rangle$, and
$v=\sqrt2\langle H_2^0\rangle$. In this case,  Fig. \ref{seesawdoub}
gives the $f^2$ independent neutrino mass
\begin{align}
m^\nu_{IJ}&=(vf_{IK})(M^{-1})_{KP}(m_{PQ})
(M^{-1})_{QL}(vf_{LJ})\nonumber\\
&=\frac{v^2}{2}\frac{V_1}{V_2^2}~f^{(lN)}_{IK}
(f^{(Nn)})^{-1}_{KP}f^{(nn)}_{PQ}(f^{(Nn)})^{-1}_{QL}
f^{(lN)}_{LJ}\nonumber\\
&=\frac{v^2V_1}{2V_2^2}f^{(nn)}_{IJ}\label{NeuMass}
\end{align}
where we used (\ref{FamSym}).

 %%%%%%%%%%%%%%%%%%%%%%%%%%%%%%%%%%%%%%%%%%%%%%%%%%%%%%%%%%%%%%%%%%%%%%%%
 %%%%%%%%%%%%%%%%%%%%%%%%%%%%%%%%%%%%%%%%%%%%%%%%%%%%%%%%%%%%%%%%%%%%%%%%
\section{Some properties of $S_3$}

If the family symmetry applied to the up-type and down-type quarks
are identical, the CKM matrix would be diagonal. Therefore, it is
necessary that the family symmetry is spontaneously broken
differently for the up-type and down-type quark sectors. Let us
briefly review how the $S_3$ symmetry can be broken differently for
the up and down type quarks. The same strategy is applied to the
leptons also. There is a long list of references on $S_3$, some of
which are given in \cite{S3perm,MaSi04}.

%%%%%%%%%%%%%%%%%%%%%%%%%%%%%%%%%%%%%%%%%%%%%%%%%%%%%%%%%%%%%%%%%%%%%%%%%%%
\subsection{Representations }

$S_3$ is a permutation symmetry of three objects, which can be
conveniently represented as permutations of three vertical points of
equilateral triangle, ${\bf A}\sim (1,0), {\bf B}\sim
(-\frac12,\frac{\sqrt3}{2})$ and ${\bf C}\sim
(-\frac12,-\frac{\sqrt3}{2})$. In the complexified coordinate,
$(x+iy,x-iy)$, these points are represented as
\begin{equation}
{\bf A}\sim \left(\begin{array}{c} 1\\ 1\end{array}\right),\
{\bf B}\sim \left(\begin{array}{c} \omega\\
\omega^2\end{array}\right),\
{\bf C}\sim \left(\begin{array}{c} \omega^2\\
\omega\end{array}\right)
\end{equation}
where $\omega$ is a cube root of unity,
$\omega=-\frac12+i\frac{\sqrt3}{2}$.
 The permutation operation of three objects is
\begin{equation}
\begin{array}{ccc}
1\quad & 2\quad  & 3\quad \\
\downarrow\quad & \downarrow\quad  & \downarrow\quad \\
i\quad & j\quad  & k\quad
\end{array}\label{S3op1}
\end{equation}
where $\{ijk\}$ is a permutation of $\{123\}$. The operation
(\ref{S3op1}) is simply written as $(ijk)$. Then, the six operations
of $S_3$ are represented as
\begin{align}
(123)\sim \left(\begin{array}{cc} 1&0\\0& 1\end{array}\right),\
 (231)\sim \left(\begin{array}{cc} \omega&0\\0&
\omega^2\end{array}\right),\
 (312)\sim \left(\begin{array}{cc}
\omega^2&0\\0& \omega\end{array}\right)\\
(132)\sim \left(\begin{array}{cc} 0&1\\1& 0\end{array}\right),\
 (321)\sim \left(\begin{array}{cc} 0&\omega^2\\
\omega&0\end{array}\right),\
 (213)\sim \left(\begin{array}{cc}
 0& \omega\\ \omega^2 &0\end{array}\right)
\end{align}
From three objects {\bf A,B,C}, we can construct a singlet ${\bf
S}\sim(\bf A+B+C)$. The other remaining combinations form a doublet
with components ${\bf D}^\uparrow\sim ({\bf A+\omega^2 B+\omega C})$
and ${\bf D}^\downarrow\sim ({\bf A+\omega B+\omega^2 C})$.
Explicitly, we can show that
$$
{\bf D}^\uparrow\sim \left(\begin{array}{c} 1\\
0\end{array}\right),\
{\bf D}^\downarrow\sim \left(\begin{array}{c} 0\\
1\end{array}\right).
$$

%%%%%%%%%%%%%%%%%%%%%%%%%%%%%%%%%%%%%%%%%%%%%%%%%%%%%%%%%%%%%%%%%%%%%%%%%%%
\subsection{Tensor products}

Consider a tensor product from two doublets of $S_3$,
\begin{equation}
\Psi_X=\left(\begin{array}{c} \psi_X^1\\
\psi_X^2\end{array}\right),\
\Psi_Y=\left(\begin{array}{c} \psi_Y^1\\
\psi_Y^2\end{array}\right).
\end{equation}

The product representation $\Psi_X\otimes\Psi_Y$ has four elements
which have the transformation following properties under $S_3$
\begin{equation}
{\bf 1}\sim (\psi_X^1\psi_Y^2+ \psi_X^2\psi_Y^1),\ {\bf 1}^\prime
\sim (\psi_X^1\psi_Y^2- \psi_X^2\psi_Y^1),\label{TwMult1}
\end{equation}
and
\begin{equation} {\bf 2}\sim
 \left(\begin{array}{c} \psi_X^2\psi_Y^2\\ \psi_X^1\psi_Y^1
 \end{array}\right).\label{TwMult2}
\end{equation}
Repeating the multiplication rule (\ref{TwMult2}), one can construct
a singlet from three doublets of $S_3$ as
\begin{equation}
\psi_X^1\psi_Y^1\psi_Z^1\pm \psi_X^2\psi_Y^2\psi_Z^2.
\end{equation}

Including the above $2\times 2$ tensor product, a dyadic is
constructed from two triplets({\bf 3=S+D}), $ {\bf S_1+D_1}\ {\rm
and}\ {\bf S_2+D_2}$ where
\begin{align}
&{\bf 3_1}:\ \textstyle{\bf S_1}=\frac{1}{3}(f_1+f_2+f_3),\ {\bf
D_1}= \textstyle\frac{1}{3}
\left(\begin{array}{c} f_1+\omega^2 f_2+\omega f_3\\
f_1+\omega f_2+\omega^2 f_3 \end{array}\right) \label{Srep1}\\
&{\bf 3_2}:\ {\bf S_2}=
\textstyle\frac{1}{3}(f_1^\prime+f_2^\prime+f_3^\prime),\
 {\bf D_2}= \textstyle\frac{1}{3}\left(\begin{array}{c}
f_1^\prime+\omega^2 f_2^\prime+\omega f_3^\prime\\
f_1^\prime+\omega f_2^\prime+\omega^2 f_3^\prime
 \end{array}\right). \label{Srep2}
\end{align}
Let the two index $S_3$ representation be $\Phi_{ij}$ which
transforms as a diadic
\begin{equation}
\Phi_{ij}\sim \phi_i\otimes\phi_j^\prime .
\end{equation}
Similarly with (\ref{Srep1},\ref{Srep2}), we define singlets and
doublets with $\phi$s,
\begin{align}
&{\bf 1}_\phi:\
\textstyle\frac{1}{3}(\phi_1+\phi_2+\phi_3)\to\xi_1,\nonumber\\
&{\bf 2}_\phi:\ \textstyle\frac{1}{3}
 \left(\begin{array}{c}
\phi_1+\omega^2 \phi_2+\omega \phi_3\\
\phi_1+\omega \phi_2+\omega^2 \phi_3
 \end{array}\right)\to \left(\begin{array}{c}
\xi_2\\ \xi_3 \end{array}\right)\label{dyadic1}\\
&{\bf 1}_{\phi^\prime}:\ \textstyle\frac{1}{3} (\phi_1^\prime
+\phi_2^\prime +\phi_3^\prime
)\to\xi_1^\prime,\nonumber\\
&{\bf 2}_{\phi^\prime}:\ \textstyle\frac{1}{3}
 \left(\begin{array}{c}
\phi_1^\prime+\omega^2 \phi_2^\prime+\omega \phi_3^\prime\\
\phi_1^\prime+\omega \phi_2^\prime+\omega^2 \phi_3^\prime
 \end{array}\right)\to \left(\begin{array}{c}
\xi_2^\prime\\ \xi_3^\prime \end{array}\right)\label{dyadic2}
\end{align}
We can introduce $S_3$ representations having two indices
 following the transformation rules of dyadic
(\ref{dyadic1}--\ref{dyadic2}) for  two-index singlets and two-index
doublets. The nine components of the dyadic made of
(\ref{dyadic1},\ref{dyadic2}) split into the following $S_3$
multiplets, which constitute the representations of $\Phi$,
\begin{align}
&S_3=\xi_1\xi_1^\prime,\quad S_4=\xi_2\xi_3^\prime
+\xi_3\xi_2^\prime,\quad  S^\prime=\xi_2\xi_3^\prime
-\xi_3\xi_2^\prime,\\
&D_3=\xi_1 \left(\begin{array}{c}\xi_2^\prime\\ \xi_3^\prime
\end{array}\right),\
D_4= \left(\begin{array}{c} \xi_2\\ \xi_3
\end{array}\right)\xi_1^\prime,\
 D_5=\left(\begin{array}{c}\xi_3\xi_3^\prime\\ \xi_2\xi_2^\prime
\end{array}\right).
\end{align}
 Below, we will  use predominantly the dyadic symbols written with
 $\xi$s.

The renormalizable Yukawa coupling for quark masses are assumed to
arise from nonrenormalizable dimension 5 operators at the Planck
scale
\begin{equation}
\sim\frac{1}{M_{Pl}}f_if_j^\prime H\langle \Phi_{ij}\rangle
\end{equation}
where $f_i$ is the symbol for a fermion, $i,j$ are the labels of the
permutation symmetry, $H$ is a Higgs doublet which does not carry a
family index, and $\Phi_{ij}$ is the two-index scalar field. Since
$H$ is a singlet of the family group, we can consider the relevant
couplings presented in Table \ref{Couplings} with coupling constants
$\lambda$s, which are interpreted as coupling times $\langle
H^0\rangle/M_{Pl}$. There are fifteen couplings.

\begin{table}
\begin{center}
{\tiny
\begin{tabular}{c|ll}
 &  ${\bf 1}$ &   ${\bf 1}^\prime$\\
 \hline
 $\bf S_1S_2S_3$ &  $\lambda_1$($f_1$+$f_2$+$f_3$)($
 f_1^\prime$+$f_2^\prime$+$f_3^\prime$)$\xi_1\xi_1^\prime$ & \\
  $\bf S_1S_2S_4$ & $\lambda_2$($f_1$+$f_2$+$f_3$)($
 f_1^\prime$+$f_2^\prime$+$f_3^\prime$)($\xi_2\xi_3^\prime$+$
 \xi_3\xi_2^\prime$)&
  \\ $\bf S_1S_2S^\prime$ & &$\lambda_3$($f_1$+$f_2$+$f_3$)($
 f_1^\prime$+$f_2^\prime$+$f_3^\prime$)($\xi_2\xi_3^\prime$--$
 \xi_3\xi_2^\prime$) \\
 &&\\
 $\bf D_1D_2S_3$ & \parbox{6cm} {$\lambda_4$[($f_1$+$\omega^2 f_2$
 +$\omega f_3$)($f_1^\prime$+$\omega f_2^\prime$
 +$\omega^2 f_3^\prime)$\\[0.0em]
 +($f_1$+$\omega f_2$+$\omega^2 f_3$)($f_1^\prime$+$\omega^2
 f_2^\prime$
 +$\omega f_3^\prime$)]$\xi_1\xi_1^\prime$ }
 & \parbox{6cm} {$\lambda_4$[($f_1$+$\omega^2 f_2$
 +$\omega f_3$)($f_1^\prime$+$\omega f_2^\prime$
 +$\omega^2 f_3^\prime)$\\[0.0em]
 --($f_1$+$\omega f_2$+$\omega^2 f_3$)($f_1^\prime$+$\omega^2
 f_2^\prime$
+$\omega f_3^\prime$)]$\xi_1\xi_1^\prime$ }
 \\

 $\bf D_1D_2S_4$ & \parbox{6cm} {$\lambda_5$[($f_1$+$\omega^2 f_2$
 +$\omega f_3$)($f_1^\prime$+$\omega f_2^\prime$
 +$\omega^2 f_3^\prime)$\\[0.0em]
 +($f_1$+$\omega f_2$+$\omega^2 f_3$)($f_1^\prime$+$\omega^2
 f_2^\prime$+$\omega f_3^\prime$)]($\xi_2\xi_3^\prime$+
 $\xi_3\xi_2^\prime$)}  &
 \parbox{6cm} {$\lambda_5$[($f_1$+$\omega^2 f_2$
 +$\omega f_3$)($f_1^\prime$+$\omega f_2^\prime$
 +$\omega^2 f_3^\prime)$\\[0.0em]
 --($f_1$+$\omega f_2$+$\omega^2 f_3$)($f_1^\prime$+$\omega^2
 f_2^\prime$+$\omega f_3^\prime$)]($\xi_2\xi_3^\prime$+
 $\xi_3\xi_2^\prime$) }
\\

$\bf D_1D_2S^\prime$ &\parbox{6cm} {$\lambda_6$[($f_1$+$\omega^2
f_2$ +$\omega f_3$)($f_1^\prime$+$\omega f_2^\prime$
 +$\omega^2 f_3^\prime)$\\[0.0em]
 --($f_1$+$\omega f_2$+$\omega^2 f_3$)($f_1^\prime$+$\omega^2
 f_2^\prime$+$\omega f_3^\prime$)]($\xi_2\xi_3^\prime$--
 $\xi_3\xi_2^\prime$)} &\parbox{6cm} {$\lambda_6$
 [($f_1$+$\omega^2 f_2$+$\omega f_3$)($f_1^\prime$
 +$\omega f_2^\prime$+$\omega^2 f_3^\prime)$\\[0.0em]
 +($f_1$+$\omega f_2$+$\omega^2 f_3$)($f_1^\prime$+$\omega^2
 f_2^\prime$+$\omega f_3^\prime$)]($\xi_2\xi_3^\prime$--
 $\xi_3\xi_2^\prime$)}
    \\

 $\bf S_1D_2D_3$ &\parbox{6cm} {$\lambda_7$[($f_1$+$ f_2$
 +$f_3$)($f_1^\prime$+$\omega^2 f_2^\prime$
 +$\omega f_3^\prime$)$\xi_1\xi_3^\prime$\\[0.0em]
 +($f_1$+$ f_2$+$ f_3$)($f_1^\prime$+$\omega
 f_2^\prime$+$\omega^2 f_3^\prime$)$\xi_1\xi_2^\prime$]}
 &\parbox{6cm} {$\lambda_7$[($f_1$+$ f_2$
 +$f_3$)($f_1^\prime$+$\omega^2 f_2^\prime$
 +$\omega f_3^\prime$)$\xi_1\xi_3^\prime$\\[0.0em]
 --($f_1$+$ f_2$+$ f_3$)($f_1^\prime$+$\omega
 f_2^\prime$+$\omega^2 f_3^\prime$)$\xi_1\xi_2^\prime$]}
 \\

 $\bf S_1D_2D_4$ & \parbox{6cm} {$\lambda_8$[($f_1$+$ f_2$
 +$f_3$)($f_1^\prime$+$\omega^2 f_2^\prime$
 +$\omega f_3^\prime$)$\xi_3\xi_1^\prime$\\[0.0em]
 +($f_1$+$ f_2$+$ f_3$)($f_1^\prime$+$\omega
 f_2^\prime$+$\omega^2 f_3^\prime$)$\xi_2\xi_1^\prime$]}
 &\parbox{6cm} {$\lambda_8$[($f_1$+$ f_2$
 +$f_3$)($f_1^\prime$+$\omega^2 f_2^\prime$
 +$\omega f_3^\prime$)$\xi_3\xi_1^\prime$\\[0.0em]
 --($f_1$+$ f_2$+$ f_3$)($f_1^\prime$+$\omega
 f_2^\prime$+$\omega^2 f_3^\prime$)$\xi_2\xi_1^\prime$]}
 \\

 $\bf S_1D_2D_5$ &\parbox{6cm} {$\lambda_9$[($f_1$+$ f_2$
 +$f_3$)($f_1^\prime$+$\omega^2 f_2^\prime$
 +$\omega f_3^\prime$)$\xi_2\xi_2^\prime$\\[0.0em]
 +($f_1$+$ f_2$+$ f_3$)($f_1^\prime$+$\omega
 f_2^\prime$+$\omega^2 f_3^\prime$)$\xi_3\xi_3^\prime$]}
 &\parbox{6cm} {$\lambda_9$[($f_1$+$ f_2$
 +$f_3$)($f_1^\prime$+$\omega^2 f_2^\prime$
 +$\omega f_3^\prime$)$\xi_2\xi_2^\prime$\\[0.0em]
 --($f_1$+$ f_2$+$ f_3$)($f_1^\prime$+$\omega
 f_2^\prime$+$\omega^2 f_3^\prime$)$\xi_3\xi_3^\prime$]}
 \\

$\bf D_1S_2D_3$ &\parbox{6cm} {$\lambda_{10}$ [($f_1$+$\omega^2
f_2$+$\omega f_3$)($f_1^\prime$+$f_2^\prime$
 +$ f_3^\prime$)$\xi_1\xi_3^\prime$\\[0.0em]
 +($f_1$+$\omega f_2$+$\omega^2 f_3$)($f_1^\prime$+$
 f_2^\prime$+$ f_3^\prime$)$\xi_1\xi_2^\prime$]}
 &\parbox{6cm} {$\lambda_{10}$[($f_1$+$\omega^2  f_2$
 +$\omega f_3$)($f_1^\prime$+$f_2^\prime$
 +$ f_3^\prime$)$\xi_1\xi_3^\prime$\\[0.0em]
 --($f_1$+$\omega f_2$+$\omega^2  f_3$)($f_1^\prime$+$
 f_2^\prime$+$f_3^\prime$)$\xi_1\xi_2^\prime$]}
 \\

$\bf D_1S_2D_4$ &\parbox{6cm} {$\lambda_{11}$[($f_1$+$\omega^2 f_2$
+$\omega f_3$)($f_1^\prime$+$f_2^\prime$
 +$ f_3^\prime$)$\xi_3\xi_1^\prime$\\[0.0em]
 +($f_1$+$\omega f_2$+$\omega^2 f_3$)($f_1^\prime$+$
 f_2^\prime$+$ f_3^\prime$)$\xi_2\xi_1^\prime$]}
 &\parbox{6cm} {$\lambda_{11}$[($f_1$+$\omega^2  f_2$
 +$\omega f_3$)($f_1^\prime$+$f_2^\prime$
 +$ f_3^\prime$)$\xi_3\xi_1^\prime$\\[0.0em]
 --($f_1$+$\omega f_2$+$\omega^2  f_3$)($f_1^\prime$+$
 f_2^\prime$+$f_3^\prime$)$\xi_2\xi_1^\prime$]}
  \\

$\bf D_1S_2D_5$ &\parbox{6cm} {$\lambda_{12}$[($f_1$+$\omega^2  f_2$
 +$\omega f_3$)($f_1^\prime$+$f_2^\prime$
 +$ f_3^\prime$)$\xi_2\xi_2^\prime$\\[0.0em]
 +($f_1$+$\omega f_2$+$\omega^2 f_3$)($f_1^\prime$+$
 f_2^\prime$+$ f_3^\prime$)$\xi_3\xi_3^\prime$]}
 &\parbox{6cm} {$\lambda_{12}$[($f_1$+$\omega^2  f_2$
 +$\omega f_3$)($f_1^\prime$+$f_2^\prime$
 +$ f_3^\prime$)$\xi_2\xi_2^\prime$\\[0.0em]
 --($f_1$+$\omega f_2$+$\omega^2  f_3$)($f_1^\prime$+$
 f_2^\prime$+$f_3^\prime$)$\xi_3\xi_3^\prime$]}
 \\

$\bf D_1D_2D_3$ &\parbox{6cm} {$\lambda_{13}$[($f_1$+$\omega f_2$
 +$\omega^2 f_3$)($f_1^\prime$+$\omega f_2^\prime$
 +$\omega^2  f_3^\prime$)$\xi_1\xi_3^\prime$\\[0.0em]
 +($f_1$+$\omega^2 f_2$+$\omega f_3$)($f_1^\prime$+$\omega^2
 f_2^\prime$+$\omega f_3^\prime$)$\xi_1\xi_2^\prime$]}
 &\parbox{6cm} {$\lambda_{13}$[($f_1$+$\omega  f_2$
 +$\omega^2 f_3$)($f_1^\prime$+$\omega f_2^\prime$
 +$\omega^2 f_3^\prime$)$\xi_1\xi_3^\prime$\\[0.0em]
 --($f_1$+$\omega^2  f_2$+$\omega f_3$)($f_1^\prime$+$\omega^2
 f_2^\prime$+$\omega f_3^\prime$)$\xi_1\xi_2^\prime$]}

\\
$\bf D_1D_2D_4$ &\parbox{6cm} {$\lambda_{14}$[($f_1$+$\omega f_2$
 +$\omega^2 f_3$)($f_1^\prime$+$\omega f_2^\prime$
 +$\omega^2  f_3^\prime$)$\xi_3\xi_1^\prime$\\[0.0em]
 +($f_1$+$\omega^2 f_2$+$\omega f_3$)($f_1^\prime$+$\omega^2
 f_2^\prime$+$\omega f_3^\prime$)$\xi_2\xi_1^\prime$]}
 &\parbox{6cm} {$\lambda_{14}$[($f_1$+$\omega  f_2$
 +$\omega^2 f_3$)($f_1^\prime$+$\omega f_2^\prime$
 +$\omega^2 f_3^\prime$)$\xi_3\xi_1^\prime$\\[0.0em]
 --($f_1$+$\omega^2  f_2$+$\omega f_3$)($f_1^\prime$+$\omega^2
 f_2^\prime$+$\omega f_3^\prime$)$\xi_2\xi_1^\prime$]}
\\

$\bf D_1D_2D_5$ &\parbox{6cm} {$\lambda_{15}$[($f_1$+$\omega f_2$
 +$\omega^2 f_3$)($f_1^\prime$+$\omega f_2^\prime$
 +$\omega^2  f_3^\prime$)$\xi_2\xi_2^\prime$\\[0.0em]
 +($f_1$+$\omega^2 f_2$+$\omega f_3$)($f_1^\prime$+$\omega^2
 f_2^\prime$+$\omega f_3^\prime$)$\xi_3\xi_3^\prime$]}
 &\parbox{6cm} {$\lambda_{15}$[($f_1$+$\omega  f_2$
 +$\omega^2 f_3$)($f_1^\prime$+$\omega f_2^\prime$
 +$\omega^2 f_3^\prime$)$\xi_2\xi_2^\prime$\\[0.0em]
 --($f_1$+$\omega^2  f_2$+$\omega f_3$)($f_1^\prime$+$\omega^2
 f_2^\prime$+$\omega f_3^\prime$)$\xi_3\xi_3^\prime$]}
\end{tabular}
 }
\end{center}
\caption{Singlet combinations from ${\bf 3_1}={\bf S_1+D_1},{\bf
3_2}={\bf S_2+D_2}$ and $\Phi_{ij}={\bf
S_3+D_3+D_4+S_4+S^\prime+D_5}$. The overall factor $\frac13$ is
omitted.}\label{Couplings}
\end{table}

Depending on the direction of VEVs, the permutation symmetry is
broken.

If the couplings are the same
$\lambda_1=\cdots=\lambda_{15}=\lambda$, the mass matrix takes the
following form for the ${\bf 1+1}^\prime$ coupling

\begin{equation}
{\tiny {\bf 1+1}^\prime:\textstyle
\frac{\lambda}{9}\left(\begin{array}{ccc}
  {3\xi_1\xi_1^\prime+6\xi_2\xi_3^\prime
 }, & (1+2\omega)(\xi_1\xi_1^\prime +2\xi_2\xi_3^\prime),
 &(1+2\omega^2)(\xi_1\xi_1^\prime
 +2\xi_2\xi_3^\prime)\\
 +6(\xi_1\xi_3^\prime+\xi_2\xi_2^\prime+\xi_3\xi_1^\prime), &  &\\
 &&\\
 (1+2\omega^2)(\xi_1\xi_1^\prime+2\xi_2\xi_3^\prime), &
   {3\xi_1\xi_1^\prime+6\xi_2\xi_3^\prime
 } & (1+2\omega)(\xi_1\xi_1^\prime
 +2\xi_2\xi_3^\prime)\\
&+6\omega^2
(\xi_1\xi_3^\prime+\xi_2\xi_2^\prime+\xi_3\xi_1^\prime),&\\
&&\\
 (1+2\omega)(\xi_1\xi_1^\prime
 +2\xi_2\xi_3^\prime), &(1+2\omega^2)(\xi_1\xi_1^\prime
 +2\xi_2\xi_3^\prime), & {3\xi_1\xi_1^\prime+6\xi_2\xi_3^\prime
 }\\
 &&+6\omega(\xi_1\xi_3^\prime+\xi_2\xi_2^\prime+\xi_3\xi_1^\prime)
 \end{array}\right)}
\end{equation}
Then, for the vacuum direction
\begin{equation}
\langle\Phi_{11}\rangle=\langle\Phi_{22}\rangle=
\langle\Phi_{33}\rangle,
\end{equation}
we obtain a $C_3$ symmetric mass matrix,
\begin{align}
\frac{\lambda}{9}\left(\begin{array}{ccc} a&c^*&b\\[-0.2em]
b&a&c^*\\[-0.2em] c^*& b&a
\end{array}\right),
\end{align}
where
\begin{align}
a=3\xi_1\xi_1^\prime+6\xi_2\xi_3^\prime,\quad
b=(1+2\omega^2)(\xi_1\xi_1^\prime+2\xi_2\xi_3^\prime),\quad
c^*=(1+2\omega)(\xi_1\xi_1^\prime
 +2\xi_2\xi_3^\prime).\nonumber
\end{align}

This is one example how a specific direction of the permutation
group is chosen by spontaneous symmetry breaking. In gauge theories,
such idea has been extensively studied \cite{ChengLi}. For other
directions, the relations are not so simple and we do not present
them here in detail. Below, we take the viewpoint that the vacuum
chooses such directions when we assume a specific form of mass
matrix.

%%%%%%%%%%%%%%%%%%%%%%%%%%%%%%%%%%%%%%%%%%%%%%%%%%%%%%%%%%%%%%%%%%%%%%%%%%
%%%%%%%%%%%%%%%%%%%%%%%%%%%%%%%%%%%%%%%%%%%%%%%%%%%%%%%%%%%%%%%%%%%%%%%%%%
\section{Mixing matrix of light and heavy neutrinos}\label{sec:MixingLH}

To fix the MNS mixing matrix $U_{\rm MNS}$ of (\ref{UMNS}), one
needs the neutrino mixing matrix $U_\nu$ obtained from
(\ref{NeuMass}) and charged lepton mixing matrix $U_l$ obtained from
(\ref{qlYukawa}). If we identify the family symmetry ansatz even to
Majorana neutrinos, $(\ref{FamSym})=(\ref{FamSymMaj})$, the MNS
mixing matrix would be identity, leading  to no mixing angle between
different families. But the `family' structure defined for Majorana
neutrino masses can be in principle different from the family
structure defined for Dirac masses. In this spirit, let us assume
\begin{equation}
f^{(lN)}\ne f^{(nn)}
\end{equation}

\subsection{$n$-tuple maximal mixing}

The maximal mixing angle in the fit of the atmospheric neutrino data
suggests some kind of symmetry. The simple form of mass matrix with
the flavor democracy \cite{FrXing} is \begin{equation}
m\propto\left(
\begin{array}{ccc}
1\quad&1\quad&1\\ 1\quad&1\quad&1\\ 1\quad&1\quad&1
\end{array}
\right)\label{FrXing}
\end{equation}
which has one heavy and two massless neutrinos. The above flavor
democratic form belongs to a special case of permutation symmetry
$S_3$ which has been extensively studied for neutrino masses
\cite{S3perm}.

In general, the permutation symmetry of $n$ Majorana neutrinos
dictates the following type mass matrix,
\begin{equation}
m_n\propto\left(
\begin{array}{ccccc}
1\quad&r\quad&r&\cdots&r\\ r\quad&1\quad&r&\cdots&r\\
 r\quad&r\quad&1&\cdots&r\\
 \cdots&\cdots&\cdots&\cdots&\cdots \\
r\quad&r\quad&r&\cdots&1
\end{array}
\right).\label{SnSym}
\end{equation}
With a real $r$, we obtain for example the eigenvalues of $(1\pm r)$
for $n=2$, two $(1-r)$ and one $(1+2r)$ for $n=3$, and  two $(1-r)$
and $1+r(1\pm\sqrt 2 i)$ for $n=4$. Since we will be interested in
three families, we do not consider the complication arising from
$n\ge 4$. For $n=3$, there exists a hierarchy of $\Delta m^2_{ij}$,
which is useful in explaining both atmospheric and solar neutrino
data. For charged leptons, in general the mass matrix is complex and
does not take the form (\ref{SnSym}).

Diagonalizing (\ref{SnSym}), we obtain a {\it bi-maximal} unitary
transformation for the case of $n=2$ \cite{KKKK1},
 \begin{equation}
U_{\nu,2\times 2}= \left(\begin{array}{cc} \frac{1}{\sqrt2}&
\frac{1}{\sqrt2}\\-\frac{1}{\sqrt2} &
\frac{1}{\sqrt2}\end{array}\right).\label{Unutwo}
 \end{equation}
$m_2$ is diagonalized as
\begin{equation}
U^\dagger_{\nu,2\times 2} m_2U_{\nu,2\times 2}\propto
\left(\begin{array}{cc} 1-r& 0\\0 & 1+r\end{array}\right).
\end{equation}
For $n=3$, we obtain a {\it tri-maximal} (third column) unitary
transformation,
 \begin{equation}
U_{\nu,3\times 3}= \left(\begin{array}{ccc} \frac{1}{\sqrt2}&
\frac{1}{\sqrt6}&\frac{1}{\sqrt3}\\
-\frac{1}{\sqrt2} &\frac{1}{\sqrt6}&\frac{1}{\sqrt3}\\
 0 &-\frac{2}{\sqrt6}&\frac{1}{\sqrt3}\end{array}\right)
  \label{Unuthree}
 \end{equation}
which diagonalizes $m_3$,
\begin{equation}
U_{\nu,3\times 3}^\dagger m_3 U_{\nu,3\times 3}
\propto\left(\begin{array}{ccc} 1-r&0&0\\ 0&1-r&0\\
0&0&1+2r
\end{array}\right).\label{PermmassH}
\end{equation}

From (\ref{Unutwo}) and (\ref{Unuthree}), we try to make a
tri-bi-maximal matrix, with one more row and column to be added to
(\ref{Unutwo}) at our disposal. But with the form (\ref{Unuthree}),
one cannot obtain a tri-bi-maximal mixing. Using the form
(\ref{Unuthree}) directly for charged leptons is not correct anyway
since the mass matrix $m_l$ itself for charged leptons is not
Hermitian. We can consider a Hermitian matrix $m_l^\dagger m_l$ for
the diagonalization of which one uses the unitary transformation of
left-handed charged leptons $U_l$.

%%%%%%%%%%%%%%%%%%%%%%%%%%%%%%%%%%%%%%%%%%%%%%%%%%%%%%%%%%%%%%%%%%%%%%%%%
\subsection{Light neutrino mass matrix from heavy neutrino mass
matrix}

Following the scheme of the previous section, we investigate the
mass matrix $m^{(nn)}$ which is proportional to $m_\nu$. The
diagonalizing matrix of $m^{(nn)}$ will appear in the neutrino
mixing matrix. The flavor democratic form for $m^{(nn)}$ (hence also
for $m_\nu$ via the double seesaw) introduces one heavy and two
massless neutrinos. Therefore, the $S_3$ symmetric form
(\ref{SnSym}) introduces one heavy neutrino and two massive
degenerate neutrinos. If the $S_3$ symmetry is slightly broken by
$\epsilon_n$ in the degenerate subspace in the following way
\begin{align}
m^{(nn)}=c\left(\begin{array}{ccc}
 1-r&0&\epsilon_n\\
  0 &1+2r&0\\
 \epsilon_n&0&1-r \end{array}\right),
\label{MassI}
\end{align}
the degenerate mass eigenvalues are split into $(1-r\pm\epsilon_n)$
and $1+2r$. Here, the bi-maximal mixing matrix for diagonalizing
$m^{(\nu)}$ is \cite{KKKK1},
\begin{align}
U_\nu=\left(\begin{array}{ccc}
\frac{1}{\sqrt2}\quad&0\quad&-\frac{1}{\sqrt2} \\
0\quad&1\quad&0 \\
\frac{1}{\sqrt2}\quad&0\quad&\frac{1}{\sqrt2}
\end{array}\right).\label{CaseI}
\end{align}
But the form (\ref{CaseI}) does not depend on the strength of
$\epsilon_n$ \cite{KKKK1}. In fact, the mass matrix form
(\ref{MassI}) indicates that the neutrino triplet transforms as a
singlet($n_2$) plus a doublet($n_1$ and $n_3$) since the $2\times 2$
subspace of $m^{(nn)}$ has the structure of the form (\ref{MassI})
in this subspace. However, we will treat $|\epsilon_n|\ll 1$ so that
two scales of $\Delta m^2_{IJ}$ is obtained such that the
atmospheric(2-3 subspace) and solar(1-3 subspace) neutrino data are
explained.

So far we discussed the detailed structure of $f^{(nn)}$ in terms of
the $S_3$ permutation symmetry. The form (\ref{CaseI}) can arise
from a tiny breaking of the $S_3$ permutation symmetry. Note that
with $\epsilon_n=0$, we recover the $S_3$ symmetry in the original
basis (\ref{SnSym}).

%%%%%%%%%%%%%%%%%%%%%%%%%%%%%%%%%%%%%%%%%%%%%%%%%%%%%%%%%%%%%%%%%%%%%%%%%%%%
\subsection{Charged leptons}

On the other hand, the mass hierarchy of charged leptons is quite
different from neutrino masses, $m_e\ll m_\mu\ll m_\tau$. Therefore,
if  $S_3$ is a useful symmetry, the mass matrix of charged leptons
must break $S_3$ form (\ref{SnSym}) further to have three different
masses for charged leptons. Since an $S_3$ symmetric real mass
matrix form (\ref{SnSym}) has a degenerate pair which we try to
avoid, we must use a subset of $S_3$ generators, leading to
tri-maximal mixing. One obvious try is the cyclic permutation, i.e.
$\{ijk\}$ of Eq. (\ref{S3op1}) is a cyclic permutation of (123).
Thus, we choose only three elements among six $S_3$ generators,
which is a cyclic permutation in one direction, $C_3$,
\begin{equation}
P_{123}\equiv I, \quad P_{231}= \left(\begin{array}{ccc}
1\quad & 2\quad  & 3 \\[-0.3em]
\downarrow\quad & \downarrow\quad  & \downarrow \\[-0.2em]
2\quad & 3\quad  & 1
\end{array}\right), \quad
 P_{312}=\left(\begin{array}{ccc}
1\quad & 2\quad  & 3 \\[-0.3em]
\downarrow\quad & \downarrow\quad  & \downarrow \\[-0.2em]
3\quad & 1\quad  & 2
\end{array}\right).
\end{equation}
Namely, we violate the exchange symmetry among any two indices.
Choosing a subset of generators is achieved by the Higgs mechanism,
which we will explore in a future communication. Still we have a
subset permutation among {\it three} in $C_3$, there is a
possibility of tri-maximal mixing. Taking the following mass matrix
for charged leptons, consistently with the cyclic permutation $C_3$,
\begin{align}
&\ m^{(l)}=\left(\begin{array}{ccc} a&c^*&b\\[-0.2em]
b&a&c^*\\[-0.2em] c^*& b&a
\end{array}\right),\label{Masslep}
\end{align}
we have
\begin{align}
M^2_l\equiv  (m^{(l)})^\dagger m^{(l)}
=\left(\begin{array}{ccc} A&B^*&B\\[-0.2em]
B&A&B^*\\[-0.2em] B^*& B&A
\end{array}\right),\label{Mass2Lep}
\end{align}
where
$$
A=|a|^2+|b|^2+|c|^2,\quad B=a^*b+b^*c^*+ac.
$$
Indeed, the diagonalizing matrix $U_l$ turns out to be tri-maximal
\cite{HPS02},
\begin{align}
U_l=\left(\begin{array}{ccc} \frsqoth&\frsqo&\frsqoo\\[0.2em]
\frsqoth&\frsqoth&\frsqoth\\[0.2em] \frsqoth& \frsqoo&\frsqo
\end{array}\right),\label{Uchlep}
\end{align}
where $\omega$ is a square root of unity,
$\omega=-\frac12+i\frac{\sqrt3}{2}$. $U_l$ diagonalizes $M^2_l$ to
\begin{align}
U_l^\dagger M^2_l U_l=
\left(\begin{array}{ccc} A+B+B^* &0&0\\[0.2em]
0&A+B\omega+B^*\omega^2&0\\[0.2em] 0& 0& A+B\omega^2+B^*\omega
\end{array}\right).\label{M2diag}
\end{align}
Thus, we identify $A,B$ and $B^*$ as
\begin{align}
A&=\textstyle\frac13(m_e^2+m_\mu^2+m_\tau^2)\\
B&=\textstyle\frac13(m_e^2+m_\mu^2\omega^2+m_\tau^2\omega)\\
B^*&=\textstyle\frac13(m_e^2+m_\mu^2\omega+m_\tau^2\omega^2)
\end{align}

%%%%%%%%%%%%%%%%%%%%%%%%%%%%%%%%%%%%%%%%%%%%%%%%%%%%%%%%%%%%%%%%%%%%%%%%%%
\subsection{Tri-bi-maximal mixing}
Now, the MNS mixing matrix can be expressed in terms of $U_\nu$ and
$U_l$,
\begin{equation}
U_{\rm MNS}=U^\dagger_l U_\nu \ ,\ \ {\rm or}\ \ U_\nu=U_l U_{\rm
MNS}\ .\label{Master}
\end{equation}
From the neutrino mixing (\ref{CaseI}) and the charged lepton mixing
(\ref{Uchlep}), thus we obtain
\begin{equation}
U_{\rm MNS}=U_l^\dagger U_\nu= \left(\begin{array}{ccc}
\frsqoth&\frsqoth&\frsqoth\\[0.2em]
 \frsqoo&\frsqoth&\frsqo\\[0.2em]
 \frsqo &\frsqoth&\frsqoo
\end{array}\right)\left(\begin{array}{ccc}
\frac{1}{\sqrt2}&0&-\frac{1}{\sqrt2} \\[0.2em]
0&1&0 \\[-0.2em]
\frac{1}{\sqrt2}&0&\frac{1}{\sqrt2}
\end{array}\right) =\left(\begin{array}{ccc}
\frac{\sqrt2}{\sqrt3}\quad&\frsqoth\quad&0\\[0.2em]
-\frac{1}{\sqrt6}\quad&\frsqoth\quad&\frac{i}{\sqrt2} \\[0.2em]
-\frac{1}{\sqrt6}\quad&\frsqoth\quad&-\frac{i}{\sqrt2}
\end{array}\right). \label{UMNSS3}
\end{equation}
Defining  $i\nu_3$ as a new mass eigenstate, the desired
tri-bi-maximal mixing matrix results
\begin{equation}
U_{\rm MNS}=\left(\begin{array}{ccc}
\frac{\sqrt2}{\sqrt3}\quad&\frsqoth\quad&0\\[0.2em]
-\frac{1}{\sqrt6}\quad&\frsqoth\quad&\frac{1}{\sqrt2} \\[0.2em]
-\frac{1}{\sqrt6}\quad&\frsqoth\quad&-\frac{1}{\sqrt2}
\end{array}\right).\label{UMNSno}
\end{equation}

%%%%%%%%%%%%%%%%%%%%%%%%%%%%%%%%%%%%%%%%%%%%%%%%%%%%%%%%%%%%%%%%%%%%%%
%%%%%%%%%%%%%%%%%%%%%%%%%%%%%%%%%%%%%%%%%%%%%%%%%%%%%%%%%%%%%%%%%%%%%%
\section{Relation $\theta_{\rm sol}+\theta_c\simeq
\frac{\pi}{4}$}\label{sec:CKMMNSrel}

\subsection{Up type quark masses}
The intriguing phenomenological relation $\theta_{\rm sol}^{\rm
exp}+\theta_c^{\rm exp}\simeq \frac{\pi}{4}$ can be explained only
if one relates the lepton and quark sectors, which is the basic
principle of GUTs. In the quark sector, both the up and the down
type mass matrices are complex. We observe the similarity in the
hierarchies of charged lepton masses and $\Qem=-\frac13$ quark
masses \cite{PartData}
\begin{align}
 m_e&\simeq \textstyle
\frac{1}{200} m_\mu\sim 0,\quad m_\mu\simeq \frac{1}{17} m_\tau\\
m_d&\simeq\textstyle
 \frac{1}{20} m_s\sim 0,\quad m_s\simeq\frac{1}{35}
 m_b.\label{dtype}
\end{align}
Along with charged leptons,  we propose that the complex down type
quark mass matrix is $C_3$ symmetric, leading to three hierarchical
masses of 0, $\sim\frac{1}{20}-\frac{1}{35}$ and $1$. In GUTs, the
small discrepancy between charged lepton and $\Qem=-\frac13$ quark
masses is explained in various ways, for example by introducing the
Georgi-Jarlskog type term \cite{GeoJarl}.

But for the up type quarks there is a huge hierarchy of 0,
$\frac{1}{150}$ and 1, due to the very large top quark mass
\begin{equation}
m_u\simeq\textstyle
 \frac{1}{200} m_c\sim 0,\quad m_c\simeq \frac{1}{150} m_t.
\end{equation}
Here arises a question, $\lq\lq$Should we treat the up type quark
mass matrix $m^{(u)}$ as interpreting  one heavy and two degenerate
zero masses or three nondegenerate masses?" Since any perturbation
can add a small addition, it is better to treat the up type matrix
$m^{(u)}$ as the first case, namely having  one heavy and two zero
mass eigenvalues. In addition to this phenomenological observation,
treating $\Qem=-\frac13$ quarks and  $\Qem=\frac23$ quarks
differently is required to obtain a nontrivial CKM matrix. Then, it
is of the same form as $m^{(nn)}$, but not quite because one is
complex and the other is real. The matrix $m^{(u)\dagger}m^{(u)}$ is
required to have two zero eigenvalues, which must be done with the
$S_3$ symmetry.

Here we emphasize two aspects: one that the quark mass matrix is
complex and another that $u$ quark is almost massless from the
outset. Thus we introduce a flavor democratic form or an $S_2$
symmetric form with $r=\pm 1$ in the $2\times 2$ subspace with zero
entries at the other row and column. This type of mass matrix is
\begin{equation}
m^{(u)\dagger} m^{(u)}\propto\left(
\begin{array}{ccc}
0\quad&0\quad&0\\ 0\quad&1\quad& \pm 1\\ 0\quad&\pm 1\quad&1
\end{array}
\right)\label{upmass}
\end{equation}
 For an explicit demonstration, we choose the
minus sign in Eq. (\ref{upmass}) and obtain the third eigenvalue as
$m_t$ with the following diagonalizing unitary matrix
\begin{equation}
U^{(u)}=\left(\begin{array}{ccc} 1\quad &0\quad &0\\[0.2em]
0\quad&\frac{1}{\sqrt2}\quad&-\frac{1}{\sqrt2}\\[0.2em]
0\quad&\frac{1}{\sqrt2}\quad&\frac{1}{\sqrt2}
\end{array}\right)\label{topmassU}
\end{equation}
where
$U^{(u)\dagger}m^{(u)\dagger}m^{(u)}U^{(u)}=(m^{(u)\dagger}m^{(u)})_{
\rm diag}$.

%%%%%%%%%%%%%%%%%%%%%%%%%%%%%%%%%%%%%%%%%%%%%%%%%%%%%%%%%%%%%%%%%%%%%%%%%%
\subsection{Relating MNS and CKM angles}

The MNS mixing matrix and the CKM mixing matrix are given by
\begin{equation}
 \begin{split}
U_{\rm MNS}&=U^\dagger_l U_\nu, \quad  {\rm or}\ \ U_l=U_\nu U_{\rm
MNS}^\dagger ,\\
U_{\rm CKM}&=U^{(u)\dagger} U^{(d)},\quad  {\rm or}\ \
U^{(d)}=U^{(u)} U_{\rm CKM}.
\end{split} \label{Mast}
\end{equation}

In unifying models, $U^{(d)}$ and $U_l$ are usually related,
\begin{align}
U^{(d)}&=
U_l : {\text{quark-lepton\ complementarity}}\\
 U^{(d)}&= U_l^\dagger: {\rm SU(5)\ GUT}
\end{align}
If we choose the quark-lepton complementarity relation, the MNS and
CKM angles are related by
\begin{equation}
  U_{\rm CKM}U_{\rm MNS}\simeq
 U^{(u)\dagger}U_\nu:\quad{\text{quark-lepton\ complementarity}}
 \label{qlcomp}
\end{equation}
On the other hand, the SU(5) GUT relation gives
\begin{equation}
U^{(u)} U_{\rm CKM}=U_{\rm MNS}U_\nu ^\dagger:\quad {\rm SU(5)\ GUT}
 .
\end{equation}
The SU(5) GUT relation can be studied with a specific form of
$U^{(u)}$ and/or $U_\nu$. Here, we illustrate our idea with the
quark-lepton complementarity, (\ref{qlcomp}). The LHS of
(\ref{qlcomp}) relates $\theta_c$ and $\theta_{\rm sol}$. Note that
$\sin\theta_c^{\rm exp}\simeq 0.22$ which leads to $\theta_c^{\rm
exp}\simeq 0.071\pi$, and from $\cos\theta_{\rm sol}^{\rm
th}=\frac{\sqrt2}{\sqrt3}$ we have $\theta_{\rm sol}^{\rm th}\simeq
0.196\pi$; thus $\theta_c^{\rm exp}+\theta_{\rm sol}^{\rm th}\simeq
0.267\pi$. Basically, $\theta_c$ and $\theta_{\rm sol}$ are related
to $U_{11}$ elements of $U_{\rm CKM}$ and $U_{\rm MNS}$. Thus,  we
are interested in the first row of a real form of $U_{\rm CKM}$
which are parametrized by two angles $\theta_c$ and $\varphi_q$
\begin{equation}
U_{\rm CKM}^{\rm th}\simeq \left(\begin{array}{ccc}
\cos\theta_c^{\rm th}\quad&\sin\theta_c^{\rm th}\cos\varphi_q\quad&
\sin\theta_c^{\rm th}\sin\varphi_q\\[0.2em]
\times\quad& \times\quad&\times \\[0.2em]
\times\quad& \times\quad&\times
\end{array}\right).\label{Uckmuse}
\end{equation}
In the same vein, we are interested in the first column of a real
form of $U_{\rm MNS}$ which are parametrized by two angles
$\theta_{\rm sol}$ and $\varphi_l$,
\begin{equation}
U_{\rm MNS}^{\rm th}\simeq \left(\begin{array}{ccc}
\cos\theta_{\rm sol}^{\rm th}\quad&\times_{12}\quad&0\\[0.2em]
\sin\theta_{\rm sol}^{\rm th}\cos\varphi_l\quad& \times_{22}
\quad&\times_{23} \\[0.2em]
\sin\theta_{\rm sol}^{\rm
th}\sin\varphi_l\quad&\times_{32}\quad&\times_{33}
\end{array}\right)\label{Umnsuse}
\end{equation}
from which we obtain
$$
(U_{\rm CKM}^{\rm th}U_{\rm MNS}^{\rm th})_{11}=\cos\theta_c^{\rm
th}\cos\theta_{\rm sol}^{\rm th}+\sin\theta_c^{\rm
th}\sin\theta_{\rm sol}^{\rm
th}(\cos\varphi_q\cos\varphi_l+\sin\varphi_q\sin\varphi_l).
$$
From the tri-bi-maximal form (\ref{tribi}), we identify
$\cos\varphi_l=-\frac{1}{\sqrt2}$ and
$\sin\varphi_l=-\frac{1}{\sqrt2}$, giving $\varphi_l=\frac54\pi$.
So, we obtain
$\cos\varphi_q\cos\varphi_l+\sin\varphi_q\sin\varphi_l=\cos(\varphi_l
-\varphi_q)=-\frac{1}{\sqrt2}(\cos\varphi_q+\sin\varphi_q)$.

The particle data book \cite{PartData} gives  $U_{\rm CKM\
11}=(0.9739\ \rm to\ 0.9751)$ and $U_{\rm CKM\ 13}=(0.0029\ \rm to\
0.0045)$, which gives $\varphi_q\simeq (0.0049-0.0065)\pi$. Thus, we
have
\begin{align}
\left(U_{\rm CKM}^{\rm th}\right.&\left.U_{\rm MNS}^{\rm
th}\right)_{11}=\cos(\theta_c^{\rm th} +\theta_{\rm sol}^{\rm th})
+\sin\theta_c^{\rm th}\sin\theta_{\rm sol}^{\rm
th}[1+\cos(\varphi_l-\varphi_q)] \nonumber\\
 &=\textstyle \cos(\theta_c^{\rm th} +\theta_{\rm sol}^{\rm th})
+\sin\theta_c^{\rm th}\sin\theta_{\rm sol}^{\rm
th}[1-\cos(\frac{\pi}{4}-\varphi_q)]\nonumber\\
 &\textstyle\to\cos(\theta_c^{\rm th} +\theta_{\rm sol}^{\rm th})
+\sin\theta_c^{\rm exp}\sin\theta_{\rm sol}^{\rm
exp}[1-\cos(\frac{\pi}{4}-\varphi_q)]
\nonumber\\
 &=\textstyle\cos(\theta_c^{\rm th} +\theta_{\rm sol}^{\rm th})
+\sin\theta_c^{\rm exp}\sin\theta_{\rm sol}^{\rm exp} \left[1-\left(
\begin{array}{c}
\cos(0.25-0.0049)\pi\\
\cos(0.25-0.0065)\pi \end{array}\right)\right]
 \nonumber\\
&\simeq\cos(\theta_c^{\rm th} +\theta_{\rm sol}^{\rm th})+0.034.
\label{RelLe}
\end{align}
where in the third row we used the experimental value for
$\sin\theta_c^{\rm th}\sin\theta_{\rm sol}^{\rm th}[\cdots]$ since
the replacement $\theta_c^{\rm th} \to \theta_c^{\rm exp}$ would
introduce a small extra piece due to the smallness of
$\sin\theta_c$.

On the other hand, the RHS of (\ref{qlcomp}) with (\ref{topmassU})
is
\begin{equation}
U^{(u)\dagger}U_\nu=\left(\begin{array}{ccc}
\frac{1}{\sqrt2}\quad&0\quad& -\frac{1}{\sqrt2}\\[0.3em]
\frac{1}{2}\quad&\frac{1}{\sqrt2}\quad&\frac{1}{2} \\[0.3em]
\frac{1}{2}\quad&-\frac{1}{\sqrt2}\quad&\frac{1}{2}
\end{array}\right).\label{RelRi}
\end{equation}
Then, from Eqs. (\ref{RelLe}) and (\ref{RelRi}), we obtain
\begin{equation}
\cos(\theta_c^{\rm th} +\theta_{\rm sol}^{\rm th}) \simeq\textstyle
\frac{1}{\sqrt2}\label{SumRel}
\end{equation}
where the accuracy of the sum  is 5\% in view of Eq. (\ref{RelLe}),
which is a pretty good approximation. Thus, with the quark-lepton
complementarity ansatz we obtain the following approximate relation,
\begin{equation}
\theta_c^{\rm th} +\theta_{\rm sol}^{\rm th}\simeq\textstyle
\frac{\pi}{4}.\label{finsum}
\end{equation}
Let us note that  we obtained (\ref{finsum}) with the following
understanding:
\begin{itemize}
\item[({\it a})]
The possible dependence on the heavy neutrino Yukawa couplings is
removed by the family symmetry. Here, the double seesaw mechanism is
used.
\item[({\it b})]
We used the $S_3$ symmetry categorically differently for
$\Qem=0,-1,-\frac13$, and $\frac23$ fermions.  In particular, for
the up type quarks, we use the mass matrix of the form
(\ref{upmass}), leading to one heavy and two zero masses.
\item[({\it c})]
 We obtain the relation $\theta_{c}^{\rm
th}+ \theta_{\rm sol}^{\rm th}\simeq \frac14\pi$ only approximately.
\item[({\it d})] The above relation is the one given at the GUT
scale.
\end{itemize}

%%%%%%%%%%%%%%%%%%%%%%%%%%%%%%%%%%%%%%%%%%%%%%%%%%%%%%%%%%%%%%%%%%%%%%%%
%%%%%%%%%%%%%%%%%%%%%%%%%%%%%%%%%%%%%%%%%%%%%%%%%%%%%%%%%%%%%%%%%%%%%%%%
\paragraph{Corrections}

 Running the Yukawa couplings from the GUT scale down to the
electroweak scale can change the relation (\ref{finsum})
significantly. But there exist ideas that this relation is not
renormalized very much \cite{renorm}. The relation is expected to be
renormalized by large Yukawa couplings and the QCD coupling. For the
Dirac type Yukawa couplings involving heavy leptons $N$ and $n$,
they are independent from the top quark Yukawa coupling and hence
can be taken as small values. So only the top quark Yukawa coupling
is important. To use the relation (\ref{qlcomp}), the RHS is
evaluated at the unification scale and the LHS uses the experimental
values at the electroweak scale. So we do not worry about the
renormalization of the RHS.

The QCD coupling is flavor blind and hence the correction to quark
masses is universal, leading to a factor of 3 modification
\cite{BEGN} to quark masses down to 5 GeV, except that for top
quark. For top quark, the difference from 3 is
$\ln(5/175)/\ln(5/10^{16})\sim 0.1$. But this correction is not what
we are interested in since we use $m^{(u)}$ which has two zero
eigenvalues. This structure of $m^{(u)}$ is not changed. For
$m^{(d)}$, both $s$ and $b$ quarks are renormalized by the same
factor 3, and hence we expect that $U_{\rm CKM}$ is not changed very
much by $\alpha_s$. In particular, we use only $U_{\rm CKM\ 11}$
which is close to 1 before and after the $\alpha_s$ correction
 \cite{strongcorr}.

For $m^{(d)}$, the most significant change due to the large top
quark Yukawa coupling (symbolically represented as $Y_t$) is
expected to arise from the violation of $m_l=m^{(d)}$ where
$m^{(d)}$ is corrected by large $Y_t$. Both $m_l$ and $m^{(d)}$ are
expected to take the form (\ref{Masslep}) at a quark-lepton
complementary scale.
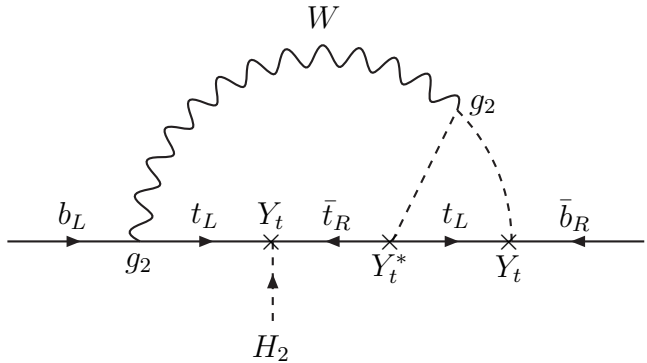
\begin{figure}[h]
\begin{center}
\begin{picture}(400,170)(0,30)
\SetWidth{0.8} \Text(200,165)[c]{$W$}
 \ArrowLine(80,80)(130,80)
\ArrowLine(320,80)(270,80)\ArrowLine(130,80)(180,80)
\ArrowLine(225,80)(180,80)\ArrowLine(225,80)(270,80)
\PhotonArc(200,80)(70,45,180){4}{12.5} \DashCArc(200,80)(70,0,45){3}

\DashLine(225,80)(248.2,127){3}

\DashArrowLine(180,50)(180,80){3}\Text(180,40)[c]{$H_2$}
\Text(155,90)[c]{$t_L$}\Text(205,90)[c]{$\bar t_R$}
\Text(225,80)[c]{$\times$}\Text(250,90)[c]{$t_L$}
\Text(180,80)[c]{$\times$}\Text(180,90)[c]{$Y_t$}
\Text(130,72)[c]{$g_2$} \Text(255,132)[l]{$g_2$}

\Text(105,90)[c]{$b_L$}\Text(295,90)[c]{$\bar b_R$}

\Text(225,70)[c]{$Y_t^*$} \Text(270,80)[c]{$\times$}
 \Text(270,70)[c]{$Y_t$}

\end{picture}
\caption{A schematic view of corrections of
$m^{(d)}$.}\label{Feynman}
\end{center}
\end{figure}
The extra correction to $m^{(d)}$ due to $Y_t$ is expected to arise
from the diagrams of the form given in Fig. \ref{Feynman}, whose
 strength is estimated roughly as
\begin{equation}
\frac{|Y_t|^3g_2^2}{(4\pi^2)^2}\cdot({\rm kinematic\ factor})\sim
\frac{|Y_t|^3\alpha_2}{4\pi^3}\sim 2.5\times 10^{-4}
\end{equation}
which is smaller than $\frac{1}{20}$ of Eq. (\ref{dtype}). Hence the
down type mass category is not drastically changed, and hence the
CKM angles are not changed drastically.

Most studies on the correction of MNS angles in GUTs have been
performed by studying the running of neutrino masses arising through
the dimension-5 operators $l lH_2H_2$ \cite{renorm}. Here, one
usually assumes a large Yukawa coupling in view of the large top
mass. Then, for non-hierarchical neutrino masses the MNS angle is
known to go a drastic change, and for hierarchical neutrino masses
the correction remains negligible. But in our double seesaw, the
needed Yukawa couplings $f_{IJ}^{(lN)}$ and $f_{IJ}^{(Nn)}$ (viz.
Eq. (\ref{NeuMass})) can be taken to be small, and running of the
MNS angle can be made negligible by taking $|f_{IJ}^{(lN)}|\ll 1$.

Therefore, we expect that the LHS of (\ref{qlcomp}) is not corrected
very much by going to the electroweak scale, and hence the sum
(\ref{finsum}) is still valid.

%%%%%%%%%%%%%%%%%%%%%%%%%%%%%%%%%%%%%%%%%%%%%%%%%%%%%%%%%%%%%%%%%%%%%%%%%%%
%%%%%%%%%%%%%%%%%%%%%%%%%%%%%%%%%%%%%%%%%%%%%%%%%%%%%%%%%%%%%%%%%%%%%%%%%%%
\paragraph{$Z_3$ orbifolds}

The $S_3$ symmetry we discuss is expected to arise from a more
fundamental theory. In the framework of quantum field theory, we can
dictate relevant couplings corresponding to the presumed family
symmetry. But it is our hope to obtain the couplings from a more
fundamental theory. One example is string theory. From string
theory, a good example allowing the permutation symmetry of 3
objects is the $Z_3$ orbifold compactification of $E_8\times
E_8^\prime$ heterotic string \cite{orbifold}. The Yukawa couplings
resulting from a $Z_3$ orbifold do not know how to distinguish the
difference of three objects, leading to the discrete symmetry. The
specific forms for Yukawa couplings are the result of spontaneous
symmetry breaking of $S_3$ symmetry. We note that in $Z_3$
orbifolds, the $\sin^2\theta_W$ problem hints toward a trinification
model \cite{trikim04}, in which we will explore a realization of the
mass matrices discussed here in a future communication.

%%%%%%%%%%%%%%%%%%%%%%%%%%%%%%%%%%%%%%%%%%%%%%%%%%%%%%%%%%%%%%%%%%%%%%%%%%
%%%%%%%%%%%%%%%%%%%%%%%%%%%%%%%%%%%%%%%%%%%%%%%%%%%%%%%%%%%%%%%%%%%%%%%%%%
\section{Conclusion}\label{sec:Conclusion}

In this paper, we use the family symmetry $S_3$ which is dictated to
be realized differently for $\Qem=0,-1,-\frac13$, and $\frac23$
fermions. We introduce two types of heavy neutral leptons $n$ and
$N$ with two additional continuous symmetries $F_1$ and $F_2$. To
discuss neutrino masses just from the symmetry principle, it is
suggested to use the double seesaw mechanism so that the Yukawa
couplings of $N$ and lepton doublets are removed. The double seesaw
diagram of Fig. \ref{seesawdoub} removes the Yukawa coupling
dependence if VEVs of singlets have a hierarchy $\langle
S_2\rangle\gg\langle S_1\rangle\gg\langle H_2\rangle$. Then one
obtains a direct proportionality between the neutrino mass matrix
and the $n$ type Majorana mass matrix, viz. Eq. (\ref{NeuMass}). Now
it becomes possible to discuss just the mass matrices, and we note
that neutrinos, charged leptons, and $\Qem=-\frac13,\frac23$ quarks
have distinct patterns of mass hierarchy. In the $S_3$ symmetric
scheme, the mass matrix forms of $n$, charged leptons and
$\Qem=-\frac13,\frac23$ quarks are dictated to be realized to
conform with the observed mass patterns. These different patterns
are the source of nontrivial MNS and CKM angles. A realization of
these different patterns is expected to result from spontaneous
symmetry breaking of family symmetry. In quark and lepton
unification models, some of these angles can be related. In this
paper, we studied the quark-lepton complementarity to relate the
charged lepton type mixing matrix and the down type quark mixing
matrix. The $SU(5)$ GUT type relation is also possible, for which
however the resulting relation is not so simple. Finally, we also
suggested a way to understand the approximate relation
$\theta_c^{\rm th} +\theta_{\rm sol}^{\rm th}\simeq \frac{\pi}{4}$.

\begin{acknowledgments}
We thank E. J. Chun, S. K. Kang and S. Raby for useful discussions.
This work is supported in part by the KRF Sundo Grant No.
R02-2004-000-10149-0, the KRF ABRL Grant No. R14-2003-012-01001-0,
and the BK21 program of Ministry of Education.
\end{acknowledgments}


\begin{thebibliography}{99}

\def\apj#1#2#3{Astrophys.\ J.\ {\bf #1} (#3) #2}
\def\ijmp#1#2#3{Int.\ J.\ Mod.\ Phys.\ {\bf #1} (#3) #2}
\def\mpl#1#2#3{Mod.\ Phys.\ Lett.\ {\bf A#1} (#3) #2}
\def\nat#1#2#3{Nature\ {\bf #1} (#3) #2}
\def\npb#1#2#3{Nucl.\ Phys.\ {\bf B#1} (#3) #2}
\def\plb#1#2#3{Phys.\ Lett.\ {\bf B#1} (#3) #2}
\def\prd#1#2#3{Phys.\ Rev.\ {\bf D#1} (#3) #2}
\def\pr#1#2#3{Phys.\ Rev.\ {\bf #1} (#3) #2}
\def\prl#1#2#3{Phys.\ Rev.\ Lett.\ {\bf #1} (#3) #2}
\def\prp#1#2#3{Phys.\ Rep.\ {\bf #1} (#3) #2}
\def\sjnp#1#2#3{Sov.\ J.\ Nucl.\ Phys.\ {\bf #1} (#3) #2}
\def\zp#1#2#3{Z.\ Phys.\ {\bf #1} (#3) #2}
\def\jhep#1#2#3{JHEP\ {\bf #1} (#3) #2}
\def\epjc#1#2#3{Euro. Phys. J.\ {\bf C#1} (#3) #2}
\def\rmp#1#2#3{Rev. Mod. Phys.\ {\bf #1} (#3) #2}
\def\prgth#1#2#3{Prog. Theor. Phys.\ {\bf #1} (#3) #2}


\bibitem{FYbook} See, for example,
M. Fukugita and T. Yanagida, {\it Physics of Neutrinos and
Application to Astrophysics}, (Springer-Verlarg, Berlin, 2003), p.
327.

\bibitem{Wolfenstein} L. Wolfenstein,
{\it Oscillations among three neutrino types and CP violation},
\prd{18}{958}{1978}.

\bibitem{HPS02} P. F. Harrison, D. H. Perkins and W. G. Scott,
{\it Tri-bimaximal mixing and the neutrino oscillation data},
\plb{530}{167}{2002} [hep-ph/0202074].

\bibitem{Ma1} E. Ma, {\it Tetrahedral family symmetry and the
neutrino mixing matrix}, [hep-ph/0508099]; {\it Pattern of the
approximate mass degeneracy of majorana neutrinos},
\mpl{17}{289}{2002}  [hep-ph/0201225]; {\it Quark mass matrices in
the $A_4$ model}, \mpl{17}{627}{2002} [hep-ph/0203238]; {\it $A_4$
symmetry and neutrinos with very different masses},
\prd{70}{031901}{2004}  [hep-ph/0404199];\\
 A. Zee, {\it Obtaining the neutrino mixing
matrix with the tetrahedral group}, \plb{630}{58}{2005}
[hep-ph/0508278];\\
 S.Luo and Z.Z. Xing, {\it Generalized tri-bimaximal neutrino
mixing and its sensitivity to radiative corrections},
 \plb{632}{341}{2006} [hep-ph/0509065].

\bibitem{bi} C. Giunti and M. Tanimoto, {\it CP violation in
bilarge lepton mixing}, \prd{66}{113006}{2002} [hep-ph/0209169];\\
P. H. Frampton, S. T. Petcov and W. Rodejohann, {\it On deviations
from bimaximal neutrino mixing}, \npb{687}{31}{2004}
[hep-ph/0401206];\\
C. H. Albright, {\it Bounds on the neutrino mixing angles
and CP phase for an $SO(10)$ model with lopsided mass matrices},
\prd{72}{013001}{2005} [hep-ph/0502161];\\
A. Datta, L. Everett and P. Ramond, {\it Cabibbo haze in lepton
mixing}, \plb{620}{42}{2005} [hep-ph/0503222];\\
S. Antusch and S. F. King, {\it Charged lepton corrections to
neutrino mixing angles and CP phases revisited}, [hep-ph/0508044];\\
F. Plentinger and W. Rodejohann, {\it Deviations from tribimaximal
neutrino mixing}, \plb{625}{264}{2005} [hep-ph/0507143].

\bibitem{complem} A. Yu Smirnov, {\it Neutrinos: $\lq$...Annus
Mirabilis'}, [hep-ph/0402264]; \\
M. Raidal, {\it Relation between the neutrino and quark mixing
angles and grand unification}, \prl{93}{161801}{2004}
[hep-ph/0404046]; \\
H. Minakata and A. Yu. Smirnov, {\it Neutrino mixing and
quark-lepton complementarity}, \prd{70}{073009}{2004}
[hep-ph/0405088]; \\
P.H. Frampton and R.N. Mohapatra, {\it Possible gauge theoretic
origin for quark-lepton complementarity}, \jhep{0501}{025}{2005}
[hep-ph/0407139]; \\
S. K. Kang, C.S. Kim and Jake Lee, {\it Importance of threshold
corrections in quark-lepton complementarity}, \plb{619}{129}{2005}
[hep-ph/0501029]; \\
J. Ferrandis and S. Pakvasa, {\it Quark-lepton complementarity
relation and neutrino mass hierarchy}, \prd{71}{033004}{2005}
[hep-ph/0412038]; \\
S. Antusch, S. F. King and R.N. Mohapatra, {\it Quark-lepton
complementarity in unified theories}, \plb{618}{150}{2005}
[hep-ph/0504007]; \\
S. F. King, {\it Predicting Neutrino parameters from $SO(3)$ family
symmetry and quark-lepton unification} [hep-ph/0506297]; \\
M. Lindner, A. Schmidt and A. Yu Smirnov, {\it Screening of Dirac
flavor structure in the seesaw and neutrino mixing},
\jhep{0507}{048}{2005}
[hep-ph/0505067]; \\
T. Ohlsson, {\it Bimaximal fermion mixing from the quark and lepton
mixing matrices}, \plb{622}{159}{2005} [hep-ph/0506094];\\
H. Minakata, {\it Quark-lepton complementarity: a review},
[hep-ph/0505262], and references therein.

\bibitem{Lindner} M. Lindener et. al., in Ref. \cite{complem}.

\bibitem{XingJarl}
Z.Z. Xing, {\it Nontrivial correlation between the CKM and MNS
matrics}, \plb{618}{141}{2005} [hep-ph/0503200];\\
   C. Jarlskog, {\it Ambiguities pertaining to quark-lepton
complementarity}, \plb{625}{63}{2005} [hep-ph/0507212].

\bibitem{BEGN} A. Buras, J. Ellis, M. K. Gaillard, and D. V.
Nanopoulos, {\it Aspects of the grand unification of strong, weak
and electromagnetic interactions}, \npb{135}{66}{1978}.

\bibitem{FrXing} H. Fritzsch and Z. Xing, {\it Lepton mass
hierarchy and neutrino oscillations}, \plb{372}{265}{1996}
[hep-ph/9509389]; {\it Large leptonic flavor mixing and the mass
spectrum of leptons}, \plb{440}{313}{1999} [hep-ph/9808272]; {\it
Maximal neutrino mixing and maximal CP violation},
\prd{61}{073016}{2000} [hep-ph/9909304];\\
E. K. Akhmedov, G. C. Branco, F. R. Joaquim, and J. I. Silva-Marcos,
{\it Neutrino masses and mixing with seesaw mechanism and universal
breaking of extended democracy}, \plb{498}{237}{2001}
[hep-ph/0008010]; \\
R. Dermisek, {\it Neutrino masses and mixing, quark-lepton symmetry
and strong right-handed neutrino hierarchy}, \prd{70}{073016}{2004}
[hep-ph/0406017]; \\
H. Fritzsch and Z. Xing, {\it Democratic
neutrino mixing reexamined}, \plb{598}{237}{2004} [hep-ph/0406206].

\bibitem{doubsee}
R. N. Mohapatra, {\it Mechanism for understanding small neutrino
mass in superstring theories}, \prl{56}{561}{1986};\\
R. N. Mohapatra and J. W. F. Valle, {\it Neutrino mass and baryon
number nonconservation in superstring models}, \prd{34}{1642}{1986};\\
For recent discussion, see for example,\\
 T. Blazek, S. Raby, and K.
Tobe, {\it Neutrino oscillations in a predictive SUSY GUT},
\prd{60}{113001}{1999}, [hep-ph/9903340]; {\it Neutrino oscillations
in an $SO(10)$ SUSY GUT with $U(2) \times U(1)^n$ family symmetry},
\prd{62}{055001}{2000} [hep-ph/9912482]; \\
R. Dermisek and S. Raby, {\it Fermion masses and neutrino
oscillations in $SO(10)$ SUSY GUT with $D(3) \times U(1)$ family
symmetry}, \prd{62}{015007}{2000} [hep-ph/9911275]; \\
S. Raby, {\it A natural framework for bilarge neutrino mixing},
\plb{561}{119}{2003} [hep-ph/0302027].

\bibitem{RossRaby} R. Barbieri, L. J. Hall, S. Raby, and A.
Romanino, {\it Unified theories with $U(2)$ flavor symmetry},
\npb{493}{3}{1997} [hep-ph/9610449].

\bibitem{Esix} F. G\"ursey, P. Ramond, and P. Sikivie, {\it
A universal gauge theory model based on $E_6$}, \plb{60}{177}{1976}.

\bibitem{kimjheptrit} S. L. Glashow, {\it Trinification of all
elementary particle forces}, in Proc. IV Workshop on Grand
Unification, ed. K. Kang et. al. (World Scientific, Singapore, 1985)
p. 88;\\
J. E. Kim, {\it SU(3) trits of orbifolded $E_8\times E_8^\prime$
heterotic string and supersymmetric standard model},
\jhep{0308}{010}{2003}  [hep-ph/0308064].

\bibitem{trikim04} J. E. Kim, {\it $Z_3$ orbifold construction of
$SU(3)^3$ GUT with $\sin^2\theta_W=\frac38$}, \plb{564}{35}{2003}
[hep-th/0301177];{\it Trinification with $\sin^2\theta_W=\frac38$
and seesaw neutrino mass}, \plb{591}{119}{2004} [hep-ph/0403196];
{\it Dynamical $\mu$ and MSSM}, \jhep{0506}{076}{2005}
[hep-th/0503031];\\
K.-S. Choi and J. E. Kim, {\it Three family $Z_3$ orbifold
trinification, MSSM, and doublet-triplet splitting problem},
\plb{567}{87}{2003} [hep-ph/0305002];\\
 K.-S. Choi, K.-Y. Choi, K. Hwang, and J. E. Kim,
{\it Higgsino mass matrix ansatz for MSSM}, \plb{579}{165}{2004}
[hep-ph/0308160].

\bibitem{ChengLi} L.-F. Li, {\it Group theory of the
spontaneously broken gauge symmetries}, \prd{9}{1723}{1974}.

\bibitem{S3perm}
 %Y. Yamaguchi, Phys. Lett. {\bf 9} (1964) 281;
S. Pakvasa and H. Sugawara, {\it Discrete symmetry and Cabibbo
angle}, \plb{73}{61}{1978};\\
G. Segr\`e and H. A. Weldon, {\it Natural suppression of strong P
and T violations and caculable mixing angles in $SU(2)\times
U(1)$}, \prl{42}{1191}{1979};\\
P. F. Harrison, D. H. Perkins, and W. G. Scott, {\it Threefold
maximal lepton mixing and the solar and atmospheric neutrino
deficits}, \plb{349}{137}{1995};\\
 K. Kang, J. E. Kim, and P. Ko,
{\it A simple modification of the maximal mixing scenario for three
light neutrinos}, \zp{72}{671}{1996}  [hep-ph/9503436].

\bibitem{MaSi04} For an introduction, see for example,\\
E. Ma, {\it Non-abelian discrete family symmetries of
 leptons and quarks}, hep-ph/0409075;\\
  N. G. Deshpanda,
  M. Gupta, and P. B. Pal, {\it Flavor-changing processes
and CP violation in the $S_3 \times Z_3$ model},
\prd{45}{953}{1992}.

\bibitem{KKKK1} K. Kang, S. K. Kang, J. E. Kim, and P. Ko,
{\it Almost maximally broken permutation symmetry for neutrino mass
matrix}, \mpl{12}{1175}{1997}  [hep-ph/9611369].

\bibitem{PartData} S. Eidelman et al., in {\it Particle Data Book},
% Particle data book, {\it The Cabibbo-Kobayashi-Maskawa Quark Mixing
 %Matrix},
 \plb{592}{1}{2004}.

\bibitem{GeoJarl} H. Georgi and C. Jarlskog, {\it A new lepton-quark
mass relation in a unified theory}, \plb{86}{297}{1979}.

\bibitem{renorm} H. Arason et. al., {\it Renormalization-group study
of the standard model and
its extensions: the standard model}, Phys. Rev. D46, 3945 (1992);\\
  J. A. Casas, J. R. Espinosa, A. Ibarra, and I. Navarro,  {\it
General RG equations for physical neutrino parameters and their
phenomenological implications}, Nucl. Phys. B \npb{573}{652}{2000}
[hep-ph/9910420];\\
  P. H. Chankowski and S. Pokorski,  {\it
  Quantum corrections to neutrino masses
and mixing angles}, \ijmp{A17}{575}{2002}
[hep-ph/0110249];\\
  S. Antusch, J. Kersten, M. Lindner, and M. Ratz,  {\it
  Running neutrino masses,
mixings and CP phases: analytical results and phenomenological
consequences}, \npb{674}{401}{2003} [hep-ph/0305273];\\
  S. Antusch, J. Kersten, M. Lindner, M. Ratz and M. A. Schmidt,
{\it Running neutrino mass parameters in see-saw scenarios},
\jhep{0503}{024}{2005} [hep-ph/0501272];\\
J. Ellis, A. Hektor, M. Kadastik, K. Kannike, and M. Raidal, {\it
Running of low-energy neutrino masses, mixing angles and CP
violation}, \plb{631}{32}{2005} [hep-ph/0506122];\\
 S.Luo and Z.Z. Xing, {\it Generalized tri-bimaximal neutrino
mixing and its sensitivity to radiative corrections},
\plb{632}{341}{2006} [hep-ph/0509065].

\bibitem{strongcorr}  H. Arason et. al., Ref. \cite{renorm}.

\bibitem{orbifold} L. J. Dixon, J. A. Harvey, C. Vafa, and E.
Witten, {\it Strings on orbifolds}, \npb{261}{678}{1985}; {\it
Strings on orbifolds 2}, \npb{274}{285}{1986};\\
 L. Iba\~nez, H. P. Nilles, and F. Quevedo,
 {\it Orbifolds and wilson lines}, \plb{187}{25}{1987};\\
 L.~Iba\~nez, J.~E.~Kim, H.~P.~Nilles, and F.~Quevedo,
{\it Orbifold compactifications with three families of $SU(3) \times
SU(2) \times U(1)^n$}, \plb{191}{282}{1987};\\
 A. Casas and C. Munoz, {\it Three generations of  $SU(3) \times
SU(2) \times U(1)^Y$ models from orbifolds}, \plb{214}{63}{1988}.

\end{thebibliography}
\end{document}